\PassOptionsToPackage{prologue,dvipsnames,table}{xcolor}
\documentclass[sigconf,screen]{acmart}

\usepackage{tcolorbox}
\usepackage{longtable}
\usepackage{balance}
\usepackage{etc-acm}
\AtBeginDocument{%
  }

\setcopyright{acmlicensed}

\acmISBN{978-1-4503-XXXX-X/18/06}

\begin{document}

\title[SENAI: Towards Software Engineering Native Generative Artificial Intelligence]{SENAI: Towards Software Engineering Native\\Generative Artificial Intelligence}
\author{Mootez Saad}
\orcid{0009-0008-8159-3632}
\affiliation{%
  \institution{Dalhouise University}
  \city{Halifax}
  \country{Canada}}
\email{mootez@dal.ca}

\author{Jos\'e Antonio Hern\'andez L\'opez}
\orcid{0000-0003-2439-2136}
\affiliation{%
  \institution{University of Murcia}
  \city{Murcia}
  \country{Spain}}
\email{joseantonio.hernandez6@um.es}

\author{Boqi Chen}
\orcid{0000-0002-1451-3603}
\affiliation{%
  \institution{McGill University}
  \city{Montr\'eal}
  \country{Canada}}
\email{boqi.chen@mail.mcgill.ca}

\author{Neil Ernst}
\orcid{0000-0001-5992-2366}
\affiliation{%
  \institution{University of Victoria}
  \city{Victoria}
  \country{Canada}}
\email{nernst@uvic.ca}

\author{D\'aniel Varr\'o}
\orcid{0000-0002-8790-252X}
\affiliation{%
  \institution{Link\"oping University}
  \city{Link\"oping}
  \country{Sweden}}
\email{daniel.varro@liu.se}

\author{Tushar Sharma}
\orcid{0000-0002-0538-052X}
\affiliation{%
  \institution{Dalhouise University}
  \city{Halifax}
  \country{Canada}}
\email{tushar@dal.ca}

\renewcommand{\shortauthors}{Saad et al.}
\newcommand{\se}{\textsc{se}}
\newcommand{\aim}{\textsc{ai}}

\begin{abstract}

Large Language Models have significantly advanced the field of code generation, demonstrating the ability to produce functionally correct code snippets. However, advancements in generative \aim{} for code overlook foundational Software Engineering (\se{}) principles such as modularity, and single responsibility, and concepts such as cohesion and coupling which are critical for creating maintainable, scalable, and robust software systems. These concepts are missing in pipelines that start with pre-training and end with the evaluation using benchmarks.

This vision paper argues for the integration of \se{} knowledge into \llms{} to enhance their capability to 
understand, analyze, and generate code and other \se{} artifacts following established \se{} knowledge.
The aim is to propose a new direction where \llms{} can move beyond mere functional accuracy to perform generative tasks that require adherence to \se{} principles and best practices.
In addition, given the interactive nature of these conversational models, we propose using Bloom's Taxonomy as a framework to assess the extent to which they internalize \se{} knowledge. The proposed evaluation framework offers a sound and more comprehensive evaluation technique compared to existing approaches such as linear probing.
Software engineering native generative models will not only overcome the shortcomings present in current models but also pave the way for the next generation of generative models capable of handling real-world software engineering.

\end{abstract}



\maketitle

\section{Introduction}
Large language models (\llms{}) have significantly transformed software development by enabling on-demand code generation and assistance capabilities. These models have demonstrated remarkable capabilities in automating programming tasks, providing code completions, and generating functions or classes based on textual prompts~\cite{Fan2023, Hou2024}. By learning from vast code repositories, \llms{} have become valuable tools for developers, enhancing productivity and reducing the time required to write boilerplate or routine code.

However, while these models are used within a software engineering context, they are not entirely trained on \textit{abstract} knowledge embodied in the field such as design principles. Indeed, most contemporary models are directly pre-trained on source code, sometimes including other sources such as Jupyter notebooks and GitHub pull requests~\cite{Lozhkov2024starcoder}. Furthermore, functional correctness is the main criterion used to evaluate their output.

\label{section:prog_se}
In \se{}, programming is a crucial step in creating concrete, operational systems that address specific problems, use cases, or functionalities from requirements. 
The evaluation and fitness of written code, however, go beyond mere functional correctness~\cite{ISO25010}. 
This is because code is rarely written in isolation; it must conform to many constraints, standards, and best practices that ensure its suitability within a larger context.

These constraints stem from the necessity for code to integrate seamlessly with existing systems, be understandable and maintainable by other developers, and remain adaptable to future requirements. Software systems evolve as requirements change or emerge, bugs are fixed, and new features are added; hence, ensuring conformance to these constraints, standards, and best practices is a constant endeavor.

More broadly, \se{} is a multifaceted domain, encompassing many development process steps (such as requirement elicitation, design, and testing), 
attempting to ensure various quality attributes (such as reliability, maintainability, and performance~\cite{ISO25010}),
while ensuring relevance of the software being developed by appropriate communication and collaboration mechanisms.
Current \llm{} development is mostly confined to models that autocomplete functionally correct code. This creates a rift between \se{} expectations and what these models deliver. This deficit in expectation was initially investigated by Pudari and Ernst~\cite{Pudari2023copilot}. They created a hierarchy of software abstractions inspired by Koopman’s Autonomous Vehicle Safety Hierarchy of Needs~\cite{Koopman2022}. Using such a hierarchy, they showed that Copilot~\cite{Chen2021evaluating} could clear out low levels of abstractions, \eg{} autocompleting syntactically correct code, but often struggled with deducing and suggesting the correct design patterns and architectural tactics for a given project.
Similarly, 
Cao \etal{}~\cite{Cao2024} demonstrated recent models such as WizardCoder~\cite{Luo2023wizardcoder} and DeepSeek-coder $33$B~\cite{deepseek-coder-33b-instruct} struggled in programming scenarios involving object-oriented programming concepts, notably \textit{inheritance} and \textit{encapsulation}.
Given these evidences, it is reasonable to assume that they may produce poorly structured software designs, leading to inflexible, hard-to-main systems. This could result in increased technical debt and higher maintenance costs.

In this paper, we first start by providing a brief overview of the steps involved in pre-training and evaluating large language models of code. Then, we lay a vision for how these models can be further aligned with \se{} practices. We propose the integration of higher-level design and architectural insights into the training process, with the aim to equip language models with a deeper understanding of software systems and with better reasoning capabilities. In addition, we propose the incorporation of other evaluation strategies to validate these models that extend beyond software correctness and other techniques for model understanding.

\section{The Status Quo}
This section provides background knowledge by 
presenting current practices used to pre-train and evaluate \llms{} for code generation. 

\subsection{Pre-training of LLMs (of Code)}

\subsubsection{Pre-training objectives:} 
\llms{} are typically trained on extensive text-based datasets such as \textit{The Pile}~\cite{Gao2020pile} and \textit{FineWeb}~\cite{Penedo2024fineweb}
that include source code and question-answer discussions involving code snippets.
Language models of code are trained specifically using source code \cite{Devlin2018bert}.
Pre-training objectives enable them to learn patterns, structures, and semantics of code to predict code-related properties and to generate code effectively.
Common pre-training objectives include next-token prediction, used by models such as \textsc{gpt}~\cite{Radford2018improving},
where the model predicts the next word based on previous words, modeling language in a left-to-right manner.

Models such as CodeT$5$~\cite{Wang2021codet5},
specialized for source code,
extend these objectives. It consolidates pre-training tasks tailored for code understanding and generation into a single framework. Specifically, it combines masking strategies by masking code spans and identifiers and trains the model to predict them, enhancing its understanding of code syntax and semantics. 
It also includes dual-generation tasks between code and natural language, training on code-to-text (\eg{} generating documentation from code) and text-to-code (\eg{} generating code from natural language descriptions), to bridge the gap between code and human language.

Recently, models such as CodeLlama~\cite{Roziere2023code} and StarCoder $2$~\cite{Lozhkov2024starcoder},
are pre-trained using the fill-in-the-middle (\textsc{fim}) objective. 
In this objective, the model is trained to generate a missing code segment between a given prefix and suffix. 
With this, the model becomes adept at understanding bidirectional context and generates code that integrates with surrounding code. This enhances its capability in tasks like code completion, refactoring, and debugging.



\subsubsection{Data representation:} Another important aspect during pre-training is the input representation fed to the language model. Cu\textsc{bert}~\cite{Kanade2020learning}, one of the earliest code language models, was trained directly on Python code snippets, treating the code as a plain sequence of tokens. 
Code\textsc{bert}~\cite{Feng2020codebert} improved upon this by introducing a bimodal representation where the input consists of a natural language sequence that describes the code snippet, followed by the code snippet. 
This approach aims to capture the semantic relationship between natural language and programming languages, particularly relevant for tasks such as code search and documentation generation.

However, treating code as a stream of tokens ignores its inherent syntactic and semantic structure, failing to model the relations and dependencies within it properly. 
To address this limitation, GraphCode\textsc{bert}~\cite{Guo2020graphcodebert} enriches the input representation by incorporating data flow graphs of the code alongside the token sequences. 
This enhanced input representation allows the model to comprehend the semantic nuances of the code more effectively, which enhances its performance on tasks that require advanced code understanding.

While the aforementioned models show good results in various code intelligence tasks, they still operate primarily at the function or file level. However, real-world scenarios involve complex dependencies with interactions between different parts of the codebase. 
By confining the pre-training phase to isolated code snippets, models may perform suboptimally when code generation necessitates repository-level knowledge. 
Recognizing this limitation, recent approaches, such as  StarCoder 2~\cite{Lozhkov2024starcoder} and DeepSeek-Coder~\cite{Guo2024deepseek},
have begun to include repository-level contextual information during the pre-training stage. 



\subsection{Benchmarks and Evaluation}
In this section, we cover some of the metrics used to quantify performance of \llms{} of code. Second, we list a non-comprehensive list of benchmarks used to evaluate these models systematically.

\subsubsection{Performance metrics:} The evaluation metrics used can be broadly categorized into two types: \textit{token-based} metrics and \textit{execution-based} metrics. Token-based metrics, such as \textsc{bleu}~\cite{Papineni2002}, \textsc{rouge}~\cite{Lin2004} and Code\textsc{bleu}~\cite{Ren2020codebleu}, evaluate the generated code by comparing it to reference code at the token or syntactic level. This provides a quantitative assessment of similarity in terms of syntax and structure. However, these metrics may not fully capture functional correctness or the ability of the code to execute successfully. 
Execution-based metrics, such as \textit{pass@k} and \textit{execution accuracy}, assess the functional correctness by executing the generated code and verifying the outputs against expected results, usually against test suites

\subsubsection{Benchmarks:} Several widely used benchmarks have been developed to evaluate language models of code, \textit{inter alia}, 
HumanEval~\cite{Chen2021evaluating}, 
\textsc{mbpp}~\cite{Austin2021program}, and recently, 
BigCodeBench~\cite{Zhuo2024bigcodebench}. 
HumanEval consists of 164 handcrafted Python programming problems, each providing a natural language description, a function signature, and hidden unit tests. Models are evaluated based on their ability to generate code that fulfills the problem descriptions and passes all the hidden tests, using metrics like pass@k to measure functional correctness. 
\textsc{mbpp} contains 974 crowd-sourced programming tasks designed to reflect challenges encountered by beginner programmers. 
It assesses models on both syntactic correctness and functional performance across a diverse set of basic programming tasks. 
BigCodeBench developed as part of the BigCode project, is designed for HumanEval-like function-level code generation tasks but with more complex instructions and diverse function calls to simulate solving practical problems faced in real-world scenarios.




\section{Vision: Software Engineering Grounded Language Models}
\begin{figure*}
    \centering
    \includegraphics[width=0.65\linewidth]{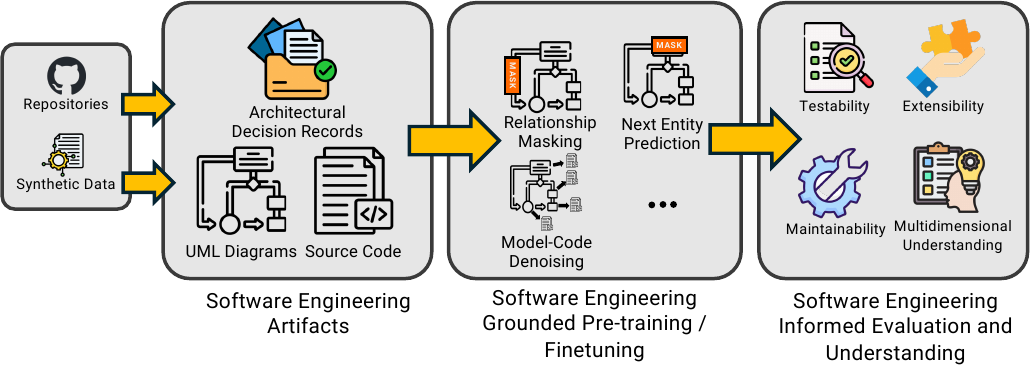}
    \caption{Overview of the vision of \se{}NAI, incorporating software engineering knowledge into the training and evaluation process of large language models.}
    \label{fig:senai-overview}
\end{figure*}
So far, \llms{} are conceived as ``(natural) language'' models, and hence natural language is the first-class citizen.
Given the textual form, these models perform code-related tasks satisfactorily.
Most of the \llm{} development, during architecture conceptualization, pre-training, or evaluation, focuses on generating syntactically and functionally correct code. However, as discussed in Section~\ref{section:prog_se}, these quality attributes are necessary, but not sufficient. In the following subsections, we lay out a vision, illustrated in Figure~\ref{fig:senai-overview}, of how to make these models \textit{grounded in \se{}}.
We discuss some of the works conducted towards this goal, their limitations, and potential new directions to address these limitations.

Native software models are grounded in \se{} and can adhere to \se{} principles and software systems.
We need \se{} native large language models because 
the existing AI models overlook the essential \se{} knowledge encapsulated in the form of principles and best practices to make software maintainable, scalable, robust, and reliable.


\subsection{Evaluation Strategy}
The evaluation process needs to be adapted to gauge the extent to which the models, old and new, grasp \se{} knowledge.

\subsubsection{Benchmarks and metrics}
\se{} principles such as modularity, cohesion, coupling, and abstraction are crucial for creating maintainable, scalable, and efficient software systems~\cite{Bass2012}. 
Assessing language models' performance concerning such principles requires a shift in both our evaluation methods and the datasets we use for assessment. Current evaluation benchmarks and metrics primarily measure the correctness of generated code against a set of test cases or reference implementations. They do not assess whether the code adheres to good software design practices. As a result, models might receive high scores even if the generated code is poorly structured or violates fundamental engineering concepts.


These new benchmarks would present tasks that require the model to demonstrate good design practices. For instance, tasks could involve, \textit{refactoring exercises}, where models are provided with low-quality code snippets (\eg{} from a maintenance perspective), and tasked to refactor them to improve quality metrics such as {\textsc{lcom}} and {\textsc{cbo}}, while maintaining functional correctness. A step further would require the model to first identify the suboptimal parts of code in a software system, and then suggest refactoring strategies that cover different levels of granularity (\ie{} from methods to packages).


\subsubsection{Model understanding}
Investigating the \se{} knowledge these models internalize is an important step towards native \se{} models.
Currently, probing methods such as linear classifier probes~\cite{Lopez2023, Ma2024}
are employed to investigate the information language models have internalized during pre-training. 
These probes analyze the model's hidden representations to determine whether specific properties or concepts are encoded in a way that can be extracted via simple classifiers. While valuable, this approach has limitations in capturing the full extent of a model's understanding, especially regarding complex and abstract \se{} concepts.

\begin{table}[h!]
\caption{Bloom's Taxonomy Levels and Corresponding Potential Assessment Questions}
\small
\rowcolors{2}{gray!15}{white}
\begin{tabular}{l|p{0.8\columnwidth}}
\rowcolor{gray!20} \textbf{Level} & \textbf{Assessment strategy} \\
\toprule
Recall & This level relates to recall from learned material. We have decided to ignore this level as it would not gauge useful information. Prompting the model on whether it recollects cohesion or coupling has limited practical implications. \\
Understand & To assess LLMs' capabilities concerning this level, the model can be prompted with code snippets, or pairs of code snippets, that exhibit different degrees (low and high) of cohesion. \\
Apply & In this stage, Code snippets with low cohesion and high coupling can be provided to the models for refactoring to improve these attributes. The software's behavior should remain the same. The evaluation will be based on test suite execution and relevant software metrics (\eg{} {\textsc{lcom}}, {\textsc{cbo}}). \\
Analyze & At the analysis level, the model is given a tuple of code fragments with varying degrees (low to high) of cohesion (or coupling) in random order. The model then ranks the snippets from lowest to highest cohesion (or coupling). Also, this level can investigate the model's understanding concerning different types of cohesion (\eg{} functional, sequential, communicational) and coupling (\eg{} temporal, data). \\
Evaluate & At this taxonomy level, the model will be given a set or pairs of classes, ranked by cohesion or a coupling measure, and asked to evaluate whether such ranking is valid. \\
Synthesis & In this stage, the model will be presented with requirements of a feature that must be implemented as a part of a software system. Then, the generated implementation will be evaluated based on correctness and its ability to produce loosely coupled and highly cohesive code. \\
\bottomrule
\end{tabular}
\label{tab:blooms_taxonomy}
\end{table}

The emergence of instruction-based language models, which allow for more flexible and interactive engagement, presents an opportunity to complement existing probing techniques with mature frameworks from psychology.
A possible framework would be Bloom's Taxonomy~\cite{Bloom1964} which categorizes cognitive skills into hierarchical levels.
This framework has been previously used by Buckley and Exton~\cite{Buckley2003} to assess developers' comprehension of a software system.
A comprehensive assessment strategy can be designed to probe the model's knowledge across various cognitive dimensions.
This framework would allow us to move beyond the surface-level analysis provided by linear probes. This includes assessing higher-order cognitive skills such as applying principles in new situations, analyzing code for adherence to best practices, evaluating design decisions, and even creating original code that embodies \se{} principles.
In addition, a taxonomy-based assessment can highlight specific areas where the model excels or underperforms. For instance, a model might effectively recall definitions but struggle with applying principles in code refactoring tasks. Identifying such gaps is crucial for guiding further pre-training and improvement. In Table~\ref{tab:blooms_taxonomy}, we present an example of a survey that can be used to examine to what extent a model internalizes foundational concepts, notably, cohesion and coupling.

\subsection{Enriching LLMs with \se{} knowledge}
To transform language models into truly \se{}-grounded tools, it is important to extend their training beyond the generation of syntactically correct and functionally accurate code. 
While these attributes are foundational, they do not encompass the full spectrum of \se{} practices that ensure code is maintainable, scalable, and aligned with sound design principles. To achieve this deeper integration, we propose several strategies on which we elaborate on blow.

\subsubsection{SE Guided Training:} 
Software development processes produce many artifacts.
Including these artifacts in the training may significantly enhance the models' capabilities.
For example, Unified Modeling Language (\textsc{uml}) diagrams provide abstract high-level representations of software systems, capturing essential aspects such as class hierarchies, object interactions, and system behaviors. 
Training models on both the code and its corresponding abstractions would enable them to understand the relationships between high-level design and concrete implementations. This multimodal learning strategy can potentially allow models to grasp architectural patterns and design principles that are not readily apparent from code alone. It can bridge the gap between conceptual understanding and practical application.

Another avenue to reinforce \se{} knowledge in the models is modifying pre-training objectives to include the prediction of entity relationships within code. By training models to recognize and predict relationships such as composition and association among classes and objects, we enhance their understanding of design and architectural structures. Similar to how models such as CodeT$5$ predict identifiers and their types, and GraphCode\textsc{bert} predicts data flow edges, incorporating entity relation prediction tasks helps models internalize the structural and semantic aspects of code. This enhancement can lead to code generation that not only functions correctly but also adheres to sound architectural practices such as promoting better cohesion and reducing coupling within the codebase.
Reinforcement Learning (\textsc{rl})-based techniques can further align code generation models. Incorporating \textsc{rl} guides the models to learn from feedback grounded in \se{} knowledge.
A work towards this direction by Weyssow \etal{}~\cite{Weyssow2024} integrates RL into code generation models to better align them with respect to code preferences, such as code readability and style.

\subsubsection{Beyond Code and Its Abstractions:} Incorporating other types of software artifacts, such as Architectural Decision Records (\textsc{adr}s), can also equip language models with reasoning abilities regarding design choices. \textsc{adr}s document the rationale behind architectural decisions, capturing the context, alternatives considered, and implications of each choice. When adding \textsc{adr}s to the training data, models might learn to associate code implementations with the underlying design intentions. This insight enables models to generate code that aligns with specific architectural tactics and to provide explanations for their design decisions, mirroring the reasoning process of experienced software engineers.

\subsubsection{Potential Challenges:} We acknowledge that these approaches present challenges, particularly concerning data procurement. Artifacts like \textsc{uml} diagrams and \textsc{adr}s are not as widely available as raw code, which could limit the volume and diversity of training data. However, there are viable strategies to mitigate this issue. Existing works in architectural recovery provide methodologies for extracting architectural information from codebases, effectively generating the necessary abstractions. Additionally, synthetic data generation techniques, such as those employed in models like WizardCoder~\cite{Luo2023wizardcoder}, can augment the training dataset. 

It is important to emphasize that we are not advocating for the abandonment of current data and pre-training methodologies. Instead, we propose enriching existing models with \se{}-grounded data and training objectives where the signal of software engineering knowledge is more \textit{amplified}. These additional signals that emphasize design principles and architectural considerations, can enhance the models' capabilities without compromising their existing strengths in code generation. This approach ensures that models retain their proficiency in generating syntactically correct and functional code, while also evolving to produce code that embodies best practices in software design.

\section{Conclusions}
Grounding \llms{} in SE principles is essential for advancing code generation beyond mere functional correctness. In this work, we called for integrating high-level design insights into language models which would steer towards the generation of code that is maintainable, scalable, and aligned with best practices. This proposal bridges the gap between code autocompletion and the demands of SE, leading to AI systems that can more effectively assist developers. Ultimately, this integration holds the promise of enhancing software quality and reducing potential technical debt in real-world applications. In parallel, we have also illustrated potential improvements to the current benchmarks and evaluation practices of language models of code.

\bibliographystyle{ACM-Reference-Format}
\balance
\bibliography{references}


\begin{thebibliography}{31}


\ifx \showCODEN    \undefined \def \showCODEN     #1{\unskip}     \fi
\ifx \showDOI      \undefined \def \showDOI       #1{#1}\fi
\ifx \showISBNx    \undefined \def \showISBNx     #1{\unskip}     \fi
\ifx \showISBNxiii \undefined \def \showISBNxiii  #1{\unskip}     \fi
\ifx \showISSN     \undefined \def \showISSN      #1{\unskip}     \fi
\ifx \showLCCN     \undefined \def \showLCCN      #1{\unskip}     \fi
\ifx \shownote     \undefined \def \shownote      #1{#1}          \fi
\ifx \showarticletitle \undefined \def \showarticletitle #1{#1}   \fi
\ifx \showURL      \undefined \def \showURL       {\relax}        \fi
\providecommand\bibfield[2]{#2}
\providecommand\bibinfo[2]{#2}
\providecommand\natexlab[1]{#1}
\providecommand\showeprint[2][]{arXiv:#2}

\bibitem[AI(2023)]%
        {deepseek-coder-33b-instruct}
\bibfield{author}{\bibinfo{person}{DeepSeek AI}.} \bibinfo{year}{2023}\natexlab{}.
\newblock \bibinfo{title}{DeepSeek Coder 33B Instruct}.
\newblock \bibinfo{howpublished}{\url{https://huggingface.co/deepseek-ai/deepseek-coder-33b-instruct}}.
\newblock
\newblock
\shownote{Accessed: 2025-Jan-16}.


\bibitem[Austin et~al\mbox{.}(2021)]%
        {Austin2021program}
\bibfield{author}{\bibinfo{person}{Jacob Austin}, \bibinfo{person}{Augustus Odena}, \bibinfo{person}{Maxwell Nye}, \bibinfo{person}{Maarten Bosma}, \bibinfo{person}{Henryk Michalewski}, \bibinfo{person}{David Dohan}, \bibinfo{person}{Ellen Jiang}, \bibinfo{person}{Carrie Cai}, \bibinfo{person}{Michael Terry}, \bibinfo{person}{Quoc Le}, {et~al\mbox{.}}} \bibinfo{year}{2021}\natexlab{}.
\newblock \showarticletitle{Program synthesis with large language models}.
\newblock \bibinfo{journal}{\emph{arXiv preprint arXiv:2108.07732}} (\bibinfo{year}{2021}).
\newblock


\bibitem[Bass et~al\mbox{.}(2012)]%
        {Bass2012}
\bibfield{author}{\bibinfo{person}{Len Bass}, \bibinfo{person}{Paul Clements}, {and} \bibinfo{person}{Rick Kazman}.} \bibinfo{year}{{2012}}\natexlab{}.
\newblock \bibinfo{booktitle}{\emph{Software Architecture in Practice, 3rd Edition}}.
\newblock
\urldef\tempurl%
\url{{https://insights.sei.cmu.edu/library/software-architecture-in-practice-third-edition/}}
\showURL{%
\tempurl}
\newblock
\shownote{Accessed: 2025-Jan-16}.


\bibitem[Bloom et~al\mbox{.}(1964)]%
        {Bloom1964}
\bibfield{author}{\bibinfo{person}{Benjamin~Samuel Bloom}, \bibinfo{person}{Max~D Engelhart}, \bibinfo{person}{Edward~J Furst}, \bibinfo{person}{Walker~H Hill}, {and} \bibinfo{person}{David~R Krathwohl}.} \bibinfo{year}{1964}\natexlab{}.
\newblock \bibinfo{booktitle}{\emph{Taxonomy of educational objectives}}. Vol.~\bibinfo{volume}{2}.
\newblock \bibinfo{publisher}{Longmans, Green New York}.
\newblock


\bibitem[Buckley and Exton(2003)]%
        {Buckley2003}
\bibfield{author}{\bibinfo{person}{J. Buckley} {and} \bibinfo{person}{C. Exton}.} \bibinfo{year}{2003}\natexlab{}.
\newblock \showarticletitle{Bloom's taxonomy: a framework for assessing programmers' knowledge of software systems}. In \bibinfo{booktitle}{\emph{11th IEEE International Workshop on Program Comprehension, 2003.}} \bibinfo{pages}{165--174}.
\newblock
\urldef\tempurl%
\url{https://doi.org/10.1109/WPC.2003.1199200}
\showDOI{\tempurl}


\bibitem[Cao et~al\mbox{.}(2024)]%
        {Cao2024}
\bibfield{author}{\bibinfo{person}{Jialun Cao}, \bibinfo{person}{Zhiyong Chen}, \bibinfo{person}{Jiarong Wu}, \bibinfo{person}{Shing-Chi Cheung}, {and} \bibinfo{person}{Chang Xu}.} \bibinfo{year}{2024}\natexlab{}.
\newblock \showarticletitle{JavaBench: A Benchmark of Object-Oriented Code Generation for Evaluating Large Language Models}. In \bibinfo{booktitle}{\emph{Proceedings of the 39th IEEE/ACM International Conference on Automated Software Engineering}}.
\newblock
\showISBNx{9798400712487}


\bibitem[Chen et~al\mbox{.}(2021)]%
        {Chen2021evaluating}
\bibfield{author}{\bibinfo{person}{Mark Chen}, \bibinfo{person}{Jerry Tworek}, \bibinfo{person}{Heewoo Jun}, \bibinfo{person}{Qiming Yuan}, \bibinfo{person}{Henrique Ponde De~Oliveira Pinto}, \bibinfo{person}{Jared Kaplan}, \bibinfo{person}{Harri Edwards}, \bibinfo{person}{Yuri Burda}, \bibinfo{person}{Nicholas Joseph}, \bibinfo{person}{Greg Brockman}, {et~al\mbox{.}}} \bibinfo{year}{2021}\natexlab{}.
\newblock \showarticletitle{Evaluating large language models trained on code}.
\newblock \bibinfo{journal}{\emph{arXiv preprint arXiv:2107.03374}} (\bibinfo{year}{2021}).
\newblock


\bibitem[Devlin(2018)]%
        {Devlin2018bert}
\bibfield{author}{\bibinfo{person}{Jacob Devlin}.} \bibinfo{year}{2018}\natexlab{}.
\newblock \showarticletitle{Bert: Pre-training of deep bidirectional transformers for language understanding}.
\newblock \bibinfo{journal}{\emph{arXiv preprint arXiv:1810.04805}} (\bibinfo{year}{2018}).
\newblock


\bibitem[Fan et~al\mbox{.}(2023)]%
        {Fan2023}
\bibfield{author}{\bibinfo{person}{Angela Fan}, \bibinfo{person}{Beliz Gokkaya}, \bibinfo{person}{Mark Harman}, \bibinfo{person}{Mitya Lyubarskiy}, \bibinfo{person}{Shubho Sengupta}, \bibinfo{person}{Shin Yoo}, {and} \bibinfo{person}{Jie~M. Zhang}.} \bibinfo{year}{2023}\natexlab{}.
\newblock \showarticletitle{{ Large Language Models for Software Engineering: Survey and Open Problems }}. In \bibinfo{booktitle}{\emph{2023 IEEE/ACM International Conference on Software Engineering: Future of Software Engineering (ICSE-FoSE)}}.
\newblock
\urldef\tempurl%
\url{https://doi.org/10.1109/ICSE-FoSE59343.2023.00008}
\showDOI{\tempurl}


\bibitem[Feng et~al\mbox{.}(2020)]%
        {Feng2020codebert}
\bibfield{author}{\bibinfo{person}{Zhangyin Feng}, \bibinfo{person}{Daya Guo}, \bibinfo{person}{Duyu Tang}, \bibinfo{person}{Nan Duan}, \bibinfo{person}{Xiaocheng Feng}, \bibinfo{person}{Ming Gong}, \bibinfo{person}{Linjun Shou}, \bibinfo{person}{Bing Qin}, \bibinfo{person}{Ting Liu}, \bibinfo{person}{Daxin Jiang}, {et~al\mbox{.}}} \bibinfo{year}{2020}\natexlab{}.
\newblock \showarticletitle{Codebert: A pre-trained model for programming and natural languages}.
\newblock \bibinfo{journal}{\emph{arXiv preprint arXiv:2002.08155}} (\bibinfo{year}{2020}).
\newblock


\bibitem[Gao et~al\mbox{.}(2020)]%
        {Gao2020pile}
\bibfield{author}{\bibinfo{person}{Leo Gao}, \bibinfo{person}{Stella Biderman}, \bibinfo{person}{Sid Black}, \bibinfo{person}{Laurence Golding}, \bibinfo{person}{Travis Hoppe}, \bibinfo{person}{Charles Foster}, \bibinfo{person}{Jason Phang}, \bibinfo{person}{Horace He}, \bibinfo{person}{Anish Thite}, \bibinfo{person}{Noa Nabeshima}, {et~al\mbox{.}}} \bibinfo{year}{2020}\natexlab{}.
\newblock \showarticletitle{The pile: An 800gb dataset of diverse text for language modeling}.
\newblock \bibinfo{journal}{\emph{arXiv preprint arXiv:2101.00027}} (\bibinfo{year}{2020}).
\newblock


\bibitem[Guo et~al\mbox{.}(2020)]%
        {Guo2020graphcodebert}
\bibfield{author}{\bibinfo{person}{Daya Guo}, \bibinfo{person}{Shuo Ren}, \bibinfo{person}{Shuai Lu}, \bibinfo{person}{Zhangyin Feng}, \bibinfo{person}{Duyu Tang}, \bibinfo{person}{Shujie Liu}, \bibinfo{person}{Long Zhou}, \bibinfo{person}{Nan Duan}, \bibinfo{person}{Alexey Svyatkovskiy}, \bibinfo{person}{Shengyu Fu}, {et~al\mbox{.}}} \bibinfo{year}{2020}\natexlab{}.
\newblock \showarticletitle{Graphcodebert: Pre-training code representations with data flow}.
\newblock \bibinfo{journal}{\emph{arXiv preprint arXiv:2009.08366}} (\bibinfo{year}{2020}).
\newblock


\bibitem[Guo et~al\mbox{.}(2024)]%
        {Guo2024deepseek}
\bibfield{author}{\bibinfo{person}{Daya Guo}, \bibinfo{person}{Qihao Zhu}, \bibinfo{person}{Dejian Yang}, \bibinfo{person}{Zhenda Xie}, \bibinfo{person}{Kai Dong}, \bibinfo{person}{Wentao Zhang}, \bibinfo{person}{Guanting Chen}, \bibinfo{person}{Xiao Bi}, \bibinfo{person}{Yu Wu}, \bibinfo{person}{YK Li}, {et~al\mbox{.}}} \bibinfo{year}{2024}\natexlab{}.
\newblock \showarticletitle{DeepSeek-Coder: When the Large Language Model Meets Programming--The Rise of Code Intelligence}.
\newblock \bibinfo{journal}{\emph{arXiv preprint arXiv:2401.14196}} (\bibinfo{year}{2024}).
\newblock


\bibitem[Hern\'{a}ndez~L\'{o}pez et~al\mbox{.}(2023)]%
        {Lopez2023}
\bibfield{author}{\bibinfo{person}{Jos\'{e}~Antonio Hern\'{a}ndez~L\'{o}pez}, \bibinfo{person}{Martin Weyssow}, \bibinfo{person}{Jes\'{u}s~S\'{a}nchez Cuadrado}, {and} \bibinfo{person}{Houari Sahraoui}.} \bibinfo{year}{2023}\natexlab{}.
\newblock \showarticletitle{AST-Probe: Recovering abstract syntax trees from hidden representations of pre-trained language models}. In \bibinfo{booktitle}{\emph{Proceedings of the 37th IEEE/ACM International Conference on Automated Software Engineering}}.
\newblock
\urldef\tempurl%
\url{https://doi.org/10.1145/3551349.3556900}
\showDOI{\tempurl}


\bibitem[Hou et~al\mbox{.}(2024)]%
        {Hou2024}
\bibfield{author}{\bibinfo{person}{Xinyi Hou}, \bibinfo{person}{Yanjie Zhao}, \bibinfo{person}{Yue Liu}, \bibinfo{person}{Zhou Yang}, \bibinfo{person}{Kailong Wang}, \bibinfo{person}{Li Li}, \bibinfo{person}{Xiapu Luo}, \bibinfo{person}{David Lo}, \bibinfo{person}{John Grundy}, {and} \bibinfo{person}{Haoyu Wang}.} \bibinfo{year}{2024}\natexlab{}.
\newblock \showarticletitle{Large Language Models for Software Engineering: A Systematic Literature Review}.
\newblock \bibinfo{journal}{\emph{ACM Trans. Softw. Eng. Methodol.}} (\bibinfo{year}{2024}).
\newblock
\urldef\tempurl%
\url{https://doi.org/10.1145/3695988}
\showDOI{\tempurl}


\bibitem[ISO/IEC 25010:2011(2011)]%
        {ISO25010}
ISO/IEC 25010:2011 \bibinfo{year}{2011}\natexlab{}.
\newblock \bibinfo{booktitle}{\emph{Systems and software engineering --- Systems and software Quality Requirements and Evaluation (SQuaRE) --- System and software quality models}}.
\newblock \bibinfo{type}{Standard} ISO/IEC 25010:2011. \bibinfo{institution}{International Organization for Standardization}, \bibinfo{address}{Geneva, Switzerland}.
\newblock


\bibitem[Kanade et~al\mbox{.}(2020)]%
        {Kanade2020learning}
\bibfield{author}{\bibinfo{person}{Aditya Kanade}, \bibinfo{person}{Petros Maniatis}, \bibinfo{person}{Gogul Balakrishnan}, {and} \bibinfo{person}{Kensen Shi}.} \bibinfo{year}{2020}\natexlab{}.
\newblock \showarticletitle{Learning and evaluating contextual embedding of source code}. In \bibinfo{booktitle}{\emph{International conference on machine learning}}. PMLR, \bibinfo{pages}{5110--5121}.
\newblock


\bibitem[Koopman(2022)]%
        {Koopman2022}
\bibfield{author}{\bibinfo{person}{Phil Koopman}.} \bibinfo{year}{2022}\natexlab{}.
\newblock \bibinfo{title}{Maturity Levels for Autonomous Vehicle Safety}.
\newblock
\newblock
\urldef\tempurl%
\url{https://safeautonomy.blogspot.com/2022/04/maturity-levels-for-autonomous-vehicle.html}
\showURL{%
\tempurl}
\newblock
\shownote{Safe Autonomy Blog}.


\bibitem[Lin(2004)]%
        {Lin2004}
\bibfield{author}{\bibinfo{person}{Chin-Yew Lin}.} \bibinfo{year}{2004}\natexlab{}.
\newblock \showarticletitle{{ROUGE}: A Package for Automatic Evaluation of Summaries}. In \bibinfo{booktitle}{\emph{Text Summarization Branches Out}}. \bibinfo{publisher}{Association for Computational Linguistics}, \bibinfo{address}{Barcelona, Spain}.
\newblock


\bibitem[Lozhkov et~al\mbox{.}(2024)]%
        {Lozhkov2024starcoder}
\bibfield{author}{\bibinfo{person}{Anton Lozhkov}, \bibinfo{person}{Raymond Li}, \bibinfo{person}{Loubna~Ben Allal}, \bibinfo{person}{Federico Cassano}, \bibinfo{person}{Joel Lamy-Poirier}, \bibinfo{person}{Nouamane Tazi}, \bibinfo{person}{Ao Tang}, \bibinfo{person}{Dmytro Pykhtar}, \bibinfo{person}{Jiawei Liu}, \bibinfo{person}{Yuxiang Wei}, {et~al\mbox{.}}} \bibinfo{year}{2024}\natexlab{}.
\newblock \showarticletitle{Starcoder 2 and the stack v2: The next generation}.
\newblock \bibinfo{journal}{\emph{arXiv preprint arXiv:2402.19173}} (\bibinfo{year}{2024}).
\newblock


\bibitem[Luo et~al\mbox{.}(2023)]%
        {Luo2023wizardcoder}
\bibfield{author}{\bibinfo{person}{Ziyang Luo}, \bibinfo{person}{Can Xu}, \bibinfo{person}{Pu Zhao}, \bibinfo{person}{Qingfeng Sun}, \bibinfo{person}{Xiubo Geng}, \bibinfo{person}{Wenxiang Hu}, \bibinfo{person}{Chongyang Tao}, \bibinfo{person}{Jing Ma}, \bibinfo{person}{Qingwei Lin}, {and} \bibinfo{person}{Daxin Jiang}.} \bibinfo{year}{2023}\natexlab{}.
\newblock \showarticletitle{Wizardcoder: Empowering code large language models with evol-instruct}.
\newblock \bibinfo{journal}{\emph{arXiv preprint arXiv:2306.08568}} (\bibinfo{year}{2023}).
\newblock


\bibitem[Ma et~al\mbox{.}(2024)]%
        {Ma2024}
\bibfield{author}{\bibinfo{person}{Wei Ma}, \bibinfo{person}{Shangqing Liu}, \bibinfo{person}{Mengjie Zhao}, \bibinfo{person}{Xiaofei Xie}, \bibinfo{person}{Wenhang Wang}, \bibinfo{person}{Qiang Hu}, \bibinfo{person}{Jie Zhang}, {and} \bibinfo{person}{Yang Liu}.} \bibinfo{year}{2024}\natexlab{}.
\newblock \showarticletitle{Unveiling Code Pre-Trained Models: Investigating Syntax and Semantics Capacities}.
\newblock \bibinfo{journal}{\emph{ACM Trans. Softw. Eng. Methodol.}} (\bibinfo{year}{2024}).
\newblock
\urldef\tempurl%
\url{https://doi.org/10.1145/3664606}
\showDOI{\tempurl}


\bibitem[Papineni et~al\mbox{.}(2002)]%
        {Papineni2002}
\bibfield{author}{\bibinfo{person}{Kishore Papineni}, \bibinfo{person}{Salim Roukos}, \bibinfo{person}{Todd Ward}, {and} \bibinfo{person}{Wei-Jing Zhu}.} \bibinfo{year}{2002}\natexlab{}.
\newblock \showarticletitle{BLEU: a method for automatic evaluation of machine translation}. In \bibinfo{booktitle}{\emph{Proceedings of the 40th Annual Meeting on Association for Computational Linguistics}} (Philadelphia, Pennsylvania) \emph{(\bibinfo{series}{ACL '02})}.
\newblock
\urldef\tempurl%
\url{https://doi.org/10.3115/1073083.1073135}
\showDOI{\tempurl}


\bibitem[Penedo et~al\mbox{.}(2024)]%
        {Penedo2024fineweb}
\bibfield{author}{\bibinfo{person}{Guilherme Penedo}, \bibinfo{person}{Hynek Kydl{\'\i}{\v{c}}ek}, \bibinfo{person}{Anton Lozhkov}, \bibinfo{person}{Margaret Mitchell}, \bibinfo{person}{Colin Raffel}, \bibinfo{person}{Leandro Von~Werra}, \bibinfo{person}{Thomas Wolf}, {et~al\mbox{.}}} \bibinfo{year}{2024}\natexlab{}.
\newblock \showarticletitle{The fineweb datasets: Decanting the web for the finest text data at scale}.
\newblock \bibinfo{journal}{\emph{arXiv preprint arXiv:2406.17557}} (\bibinfo{year}{2024}).
\newblock


\bibitem[Pudari and Ernst(2023)]%
        {Pudari2023copilot}
\bibfield{author}{\bibinfo{person}{Rohith Pudari} {and} \bibinfo{person}{Neil~A Ernst}.} \bibinfo{year}{2023}\natexlab{}.
\newblock \showarticletitle{From copilot to pilot: Towards AI supported software development}.
\newblock \bibinfo{journal}{\emph{arXiv preprint arXiv:2303.04142}} (\bibinfo{year}{2023}).
\newblock


\bibitem[Radford(2018)]%
        {Radford2018improving}
\bibfield{author}{\bibinfo{person}{Alec Radford}.} \bibinfo{year}{2018}\natexlab{}.
\newblock \showarticletitle{Improving language understanding by generative pre-training}.
\newblock  (\bibinfo{year}{2018}).
\newblock


\bibitem[Ren et~al\mbox{.}(2020)]%
        {Ren2020codebleu}
\bibfield{author}{\bibinfo{person}{Shuo Ren}, \bibinfo{person}{Daya Guo}, \bibinfo{person}{Shuai Lu}, \bibinfo{person}{Long Zhou}, \bibinfo{person}{Shujie Liu}, \bibinfo{person}{Duyu Tang}, \bibinfo{person}{Neel Sundaresan}, \bibinfo{person}{Ming Zhou}, \bibinfo{person}{Ambrosio Blanco}, {and} \bibinfo{person}{Shuai Ma}.} \bibinfo{year}{2020}\natexlab{}.
\newblock \showarticletitle{Codebleu: a method for automatic evaluation of code synthesis}.
\newblock  (\bibinfo{year}{2020}).
\newblock


\bibitem[Roziere et~al\mbox{.}(2023)]%
        {Roziere2023code}
\bibfield{author}{\bibinfo{person}{Baptiste Roziere}, \bibinfo{person}{Jonas Gehring}, \bibinfo{person}{Fabian Gloeckle}, \bibinfo{person}{Sten Sootla}, \bibinfo{person}{Itai Gat}, \bibinfo{person}{Xiaoqing~Ellen Tan}, \bibinfo{person}{Yossi Adi}, \bibinfo{person}{Jingyu Liu}, \bibinfo{person}{Romain Sauvestre}, \bibinfo{person}{Tal Remez}, {et~al\mbox{.}}} \bibinfo{year}{2023}\natexlab{}.
\newblock \showarticletitle{Code llama: Open foundation models for code}.
\newblock \bibinfo{journal}{\emph{arXiv preprint arXiv:2308.12950}} (\bibinfo{year}{2023}).
\newblock


\bibitem[Wang et~al\mbox{.}(2021)]%
        {Wang2021codet5}
\bibfield{author}{\bibinfo{person}{Yue Wang}, \bibinfo{person}{Weishi Wang}, \bibinfo{person}{Shafiq Joty}, {and} \bibinfo{person}{Steven~CH Hoi}.} \bibinfo{year}{2021}\natexlab{}.
\newblock \showarticletitle{Codet5: Identifier-aware unified pre-trained encoder-decoder models for code understanding and generation}.
\newblock \bibinfo{journal}{\emph{arXiv preprint arXiv:2109.00859}} (\bibinfo{year}{2021}).
\newblock


\bibitem[Weyssow et~al\mbox{.}(2024)]%
        {Weyssow2024}
\bibfield{author}{\bibinfo{person}{Martin Weyssow}, \bibinfo{person}{Aton Kamanda}, {and} \bibinfo{person}{Houari Sahraoui}.} \bibinfo{year}{2024}\natexlab{}.
\newblock \showarticletitle{CodeUltraFeedback: An LLM-as-a-Judge Dataset for Aligning Large Language Models to Coding Preferences}.
\newblock \bibinfo{journal}{\emph{arXiv preprint arXiv:2403.09032}} (\bibinfo{year}{2024}).
\newblock


\bibitem[Zhuo et~al\mbox{.}(2024)]%
        {Zhuo2024bigcodebench}
\bibfield{author}{\bibinfo{person}{Terry~Yue Zhuo}, \bibinfo{person}{Minh~Chien Vu}, \bibinfo{person}{Jenny Chim}, \bibinfo{person}{Han Hu}, \bibinfo{person}{Wenhao Yu}, \bibinfo{person}{Ratnadira Widyasari}, \bibinfo{person}{Imam Nur~Bani Yusuf}, \bibinfo{person}{Haolan Zhan}, \bibinfo{person}{Junda He}, \bibinfo{person}{Indraneil Paul}, {et~al\mbox{.}}} \bibinfo{year}{2024}\natexlab{}.
\newblock \showarticletitle{Bigcodebench: Benchmarking code generation with diverse function calls and complex instructions}.
\newblock \bibinfo{journal}{\emph{arXiv preprint arXiv:2406.15877}} (\bibinfo{year}{2024}).
\newblock


\end{thebibliography}

\end{document}